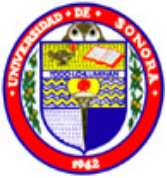

Departamento de Física
Universidad de Sonora

# A New Structural parameter for Perovskite Type Compounds and Its Application to 'Stripe Phases'

Raúl Pérez-Enríquez
rpereze@fisica.uson.mx

**Abstract.** A new index useful for the study of the Perovskite type Superconductors has been introduced. The index named Structural Parameter of High Tc Superconductivity (PESATc from its name in Spanish) uses the geometrical properties of this kind of ceramic compound. PESATc index's calculation requires the knowledge of atomic positions in the structure that could be measured by neutron diffraction or x-ray diffraction. The aim of this index is to relate the structural cell parameters with the length of Cu-O(4) bond, through an orbital specially defined: the Möbius Orbital. This is defined as the length of an Octahedral Möbius Strip overlaying the elongated octahedron created by the CuO planes and the mentioned apical Oxygen atoms, O(4). It has been found that critical temperature correlates linearly with PESATc for a set of Rare Earth base compounds (RBCO:123); the slope of the curve obtained this way has a value of 4.515 with correlation coefficient of 0.9402. Due to the alignment of the octahedral structure along the crystal, Möbius Orbital could give a clue to the understanding 'stripe phases' (phase separation in systems of spin and charge) observed in another kind of perovskite type compounds: (La,Ca)MnO$_3$, for example.

Blackett P, "Elementary Topology: A Combinatorial and Algebraic Approach", *Academic Press* (1967).
Kivelson S., "Stripe Phases in High Temperature Superconductors, *Stanford HTCS Colloquium* () (1999)
Mori C.H., "Pairing of charge-ordered stripes in (La,Ca)MnO3, *Nature* **392**, (1998) p. 473-476.
Pérez-Enríquez R, "A Structural Parameter for High Tc Superconductivity from an Octahedral Möbius Strip in RBaCuO:123 type Perovskites", RMF, suplemento en prensa (2001).

Rev Mex Fis v.48 supplement 1, sept (2002)

## The Octahedral Representation of Möbius Strip

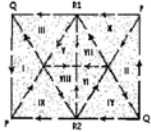

- The Möbius Strip is a surface with only one side and only one face
- Must relevant property is its non-orientability that makes it seem like electron spin
- A Möbius Strip can be divided by polygons to build a 3-dimensional octahedron
- Its Euler characteristic is equal to N0 – N1 + N2 = 0, with N values of 7, 17 and 10, respectively
- The Octahedral Strip could be modified in order to get the shape that could be overlaid in the Perovskite's octahedral structures
- A Octahedral Möbius Length is defined

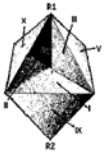

## The YBaCuO and Its Perovskite Type Structure

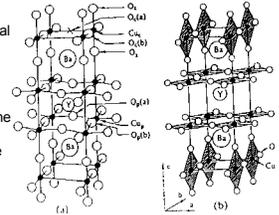

- The ideal structure of Perovskites has a general equation ABX$_3$; being A and B metallic cations, while the third element, X, is a non-metallic ion (example, Calcium Titanate (CaTiO$_3$))
- In YBa$_2$Cu$_3$O$_{7-y}$, ideal structure of Perovskite type changes due to the presence of Y ions at the Ba sites
- As the Oxygen occupancy (7-y) increases, the structure presents CuO layers and CuO chains
- The planes form octahedral structures when apical Oxygen atoms, O(4), are included.

## P E S A T c .- Structural Parameter of High Tc Superconductivity

- In order to obtain PESATc index, an Octahedral Möbius Strip overlaying the Perovskite structure of YbaCuO is required
- Observed features that support the PESATc concept:
  - a) There is a giant anisotropy of resistivity $\rho_c/\rho_{ab}$. In the limit as t->0, $\rho_c$->00 (no coherent transport on c-direction);
  - b) the new two-dimensional state (Luttinger liquid) implies a separation between charge and spin excitations for the electron;
  - c) the major contributor to superconducting condensation energy is the interlayer hopping together with the above mentioned state of electrons; and,
  - d) the non-orientable behaviour of electron spin as shown form space quantization
- The parameter involve all characteristic dimensions of the unitary cell: a, b and c parameters; as well as distance between planes
- PESATc could be useful for the analysis of other layered superconducting materials
- The name PESATc is due to *Parámetro de la Superconductividad de Alta Temperatura crítica*

$$PESATc = \frac{Octahedral\ Möbius\ Length}{Cu\text{-}O(4)\ Bond\ Length}$$

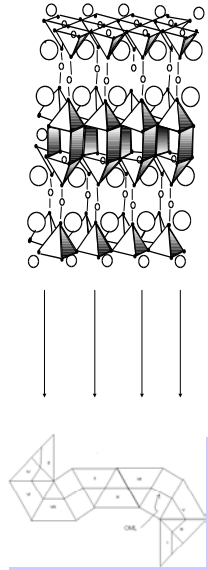

○ Ba
○ Y
○ Cu
● O

## Stripe Phases of Charge and Spin observed in Manganates and Cuprates

**Recently, the presence of stripe phases in systems of charge and spin has attracted much attention. This behaviour has been proposed as a mean to understand the physics of High Tc Superconductors. An effect of Pairing of charge-ordered stripe has been observed in certain compounds like (La,Ca)MnO3. An important fact in those manganese oxides is that they have Perovskite structure similar to that observed in copper oxide superconductors. Also, they present other extraordinary electronic and magnetic properties as the colosal magnetoresistance.**

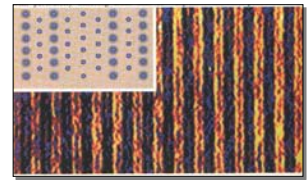

## Behaviour of Tc vs PESATc for YBaCuO Compounds

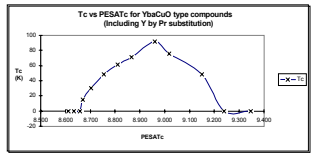

- In the graph, it is possible to see that there is a relation of Tc with PESATc:
  - a) it defines a range of values for which the compound is not a superconductor;
  - b) it has a corresponding value of 8.96 for maximum Tc in the YBaCuO compound; and,
  - c) in the case for Oxygen content samples values (A1..H1), the curve obtained shows a smooth behaviour different from that obtained when Tc is directly graphed vs the Oxygen content.

The trend of Tc vs PESATc for RBaCuO compounds becomes a straight line as can be seen:
  - a) The slope of the straight line rises to a value of 4.15;
  - b) The correlation coefficient (R = 0.904) gives a very good confidence; and,
  - c) The reason for the omission of Lanthanum (La) based compound is because it goes out of the general trend.

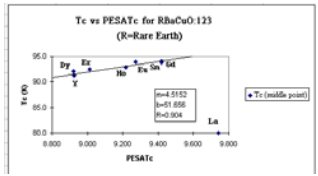

| TABLE III. PESATc* and Tc for YBaCuO type compounds | | | | | | | | | | | |
|---|---|---|---|---|---|---|---|---|---|---|---|
| Sample | A2 | B2 | C2 | D2 | E1 | A1 | B1 | C1 | D1 | E1 | F1 | G1 | H1 |
| PESATc | 9.35 | 9.24 | 9.15 | 9.02 | 8.96 | 8.87 | 8.81 | 8.75 | 8.70 | 8.67 | 8.66 | 8.65 | 8.61 |
| Tc | 0 | 0 | 48 | 76 | 92 | 71 | 61 | 48 | 30 | 15 | 0 | 0 | 0 |

*Calculated from Neutron Diffraction data [4, 13]

## Conclusions

- **PESATc Index shows to be a useful structural parameter for YBaCuO Perovskite type superconductors**
- **For YBaCuO, the behaviour of Tc vs PESATc has a smoothly increasing trend between 8.66 and 8.9 where Tc acquires its maximum of 92 K.**
- **Similar structured compounds must be analysed in order to show the PESATc's generality**
- **Perovskite type compounds like (La,Ca)MnO3 show a Striped phase behaviour**
- **Magnetic and Lattice fluctuations in YBaCuO could be explained by Stripe phase model**
- **Stripe Phases could be related with the Octahedral Möbius Orbitals used while defining PESATc**

## Stripe Phases in LaxCa1-xMnO3

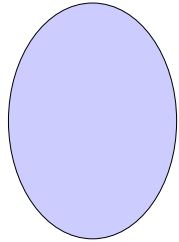

Pairing of charge-ordered stripes in La0.33Ca0.67MnO3:

**a**, High-resolution lattice image obtained at 95 K showing 3ao pairing of JTS.

**b**, Schematic modal in the **a-b** plane showing the pairing and orbital ordering of Mn3+ JTS in blue, and the Mn4+ ions in orange.

**c**, Inverted intensity scan from a selected area in **a** (a.u., arbitrary units).

Note. Pairs of the strongest peaks are identified as Mn3+O6 JTS (Octahedral) and the weaker peaks and valleys as Mn4+O6 stripes. The same horizontal distance scale is used in **b** and **c** to facilitate the comparison. (Mori S *et al*. 1998).

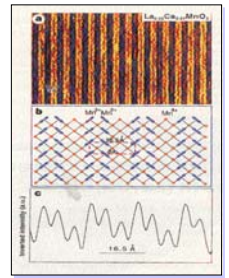

?